\documentstyle{frank}

\def\d{\,{\rm d}}
\def\bcn{\begin{center}}
\def\ecn{\end{center}}
\newcommand{\bea}{\begin{eqnarray}}
\newcommand{\eea}{\end{eqnarray}}
\newcommand{\beq}{\begin{equation}}
\newcommand{\eeq}{\end{equation}}

\newcommand{\ml}{maximum likelihood}

\newcommand{\lc}{linear collider}
\newcommand{\sm}{standard model}

\newcommand{\xs}{cross section}

\newcommand{\EM}{electromagnetic}

\newcommand{\susic}{supersymmetric}

\newcommand{\cm}{center of mass}

\newcommand{\pe}{\mbox{$e^+e^-$}}

\def\lr3{$SU(3)_L\otimes SU(3)_R$}

\def\z0{$Z^0$}
\def\Z0{$Z^0$}

\def\cm{centre of mass}
\def\gsim{\buildrel{\lower.7ex\hbox{$>$}}\over{\lower.7ex\hbox{$\sim$}}}
\def\lsim{\buildrel{\lower.7ex\hbox{$<$}}\over{\lower.7ex\hbox{$\sim$}}}
\newenvironment{comment}[1]{}{}

\begin{document}

\begin{flushright}
PSI-PR-96-11 \\
March 1996
\end{flushright}

\vskip2cm

\begin{frontmatter}
\title{The Resolving Power of a Reaction}
\author{Frank Cuypers}
\address{{\tt cuypers@pss058.psi.ch}\\
        Paul Scherrer Institute,
        CH-5232 Villigen PSI,
        Switzerland}
\begin{abstract}
We describe a simple method to compute 
the Cram\'er-Rao limit of a high energy experiment,
{\em i.e.},
the smallest error
with which a parameter can in principle be determined
in a reaction.
This precision remains a theoretical paradigm
since it assumes perfect experimental conditions.
Nevertheless,
it is shown at hand of an example
that for simple processes
this asymptotic resolving power can be approached very closely.
In all situations,
the procedure is at least a useful test
of what {\em could} and what {\em cannot}
be measured by studying a particular reaction.
\end{abstract}
\end{frontmatter}
\clearpage

\section{Introduction}

It is customary in high energy physics 
to anticipate experimental results
and to determine many years 
in advance of an experiment
how precisely it can measure a parameter.
For instance,
in the past few years a true industry has been developed
to estimate the discovery potential of LEP II.
In particular,
the reaction $\pe \to W^+W^-$ is a prime candidate
for testing anomalous gauge couplings,
since it involves the as yet unprobed 
$WW\gamma$ and $WWZ$ vertices.
Typically,
one assumes a particular form for these couplings
(generally, their \sm\ prediction)
and then proceeds to determine the expected experimental error bounds
around this central value.

In general, 
this procedure depends on four ingredients:
\begin{itemize}
\item A theory
  ({\em e.g.}, the \sm, its \susic\ extension, {\em etc.})
  which depends on one or more parameters
  (couplings, masses, {\em etc.}).
  It is the precision with which these parameters can be determined
  we wish to compute.
\item A reaction characterized by its initial and final state
  ({\em e.g.}, $\pe \to \mu^+\mu^-$ with or without polarization).
  This reaction should of course be as sensitive as possible
  to the values taken by the parameters.
\item An observable of this reaction
  ({\em e.g.}, the total \xs, asymmetries, {\em etc.}).
  It should obviously also depend as much as possible
  on the parameters.
\item A consistent, unbiased and efficient statistical estimator.
  It is generally chosen to be a least squares or maximum likelihood estimator,
  which are both equivalent and optimal in the asymptotic limit.
\end{itemize}

The issue we wish to address here
is how to optimize the last two of these four items.
For this
we shall assume a perfect experiment
with no other errors than statistical ones.
We shall introduce a {\em theoretical} observable and a statistical estimator,
which yield the smallest possible error on the parameters 
that can be obtained
with a given amount of data.
This theoretical limit
is nothing else but the Cram\'er-Rao minimum variance bound \cite{pdg}.
It clearly defines
a boundary between what in principle can be achieved
and what certainly cannot be achieved,
by studying a particular reaction.
In the experimental practice,
of course,
it remains the task of the physicist
to make use of an observable
(or a set of observables)
which yields a sensitivity
that comes close to this asymptotic resolving power.

In the next Section we define the $\chi^2_\infty$ estimator,
which computes the Cram\'er-Rao limit 
of the error in the determination of a parameter.
In Section 3,
we use this criterion 
to derive limits for an electric dipole moment of the electron
in a high energy M\o ller scattering experiment.
Because this reaction is particularly simple
it allows the derivation of analytical formulae
which nicely exhibit some general features of the procedure.
In Section 4,
we consider a similar analysis in Compton scattering.
This example will display how realistic a goal
the result of the $\chi^2_\infty$ estimator can be
when the phase space is larger.
Finally, 
we recapitulate in the Conclusion
the aim and the domain of applicability of this estimator.

\section{The Cram\'er-Rao Limit}

Let us consider a generic high-energy scattering experiment
and a theory which by assumption is the correct one.
For simplicity
we concentrate here on the determination of a single parameter $\rho$
of this theory.
It is straightforward to extend all results to follow
to the case where several parameters are involved.
The true value of the parameter is $\tilde \rho$.

We wish to determine the range of values of $\rho$
which would be indistinguishable from $\tilde \rho$
when a particular measurement is performed.
For example,
one could compare the total predicted rates $n(\tilde \rho)$ and $n(\rho)$.
The values of $\rho$ for which
\beq
n(\tilde \rho) - \chi_1 \Delta n(\tilde \rho)
\quad < \quad n(\rho) \quad < \quad
n(\tilde \rho) + \chi_1 \Delta n(\tilde \rho)
\label{e0}
\eeq
cannot be distinguished from $\tilde \rho$
to better than $\chi_1$ standard deviations.
The average numbers of events $n$ 
are computed by integrating
the differential \xs s
over the final state phase space $\Omega$
which can be explored by the experiment:
\beq
n = {\cal L} \sigma = {\cal L} \int\!\d\Omega\ {\d\sigma \over \d\Omega}
~,\label{e2}
\eeq
where $\cal L$ is the time integrated luminosity.
If systematic errors can be neglected
the numbers of events are distributed according to Poisson statistics,
and the standard deviation in Eq.~(\ref{e0}) is given by
\beq
\Delta n = \sqrt{n}
~.\label{e3}
\eeq
In order to allow an easy generalization,
we can rewrite Eqs~(\ref{e0},\ref{e3}) as a least squares estimator
\bea
\chi^2_1 
& = &
\left( {n(\rho)-n(\tilde \rho) \over \Delta n(\tilde \rho)} \right)^2
\nonumber\\
& = &
{\cal L}\
{\left[ \displaystyle\int\!\d\Omega\ 
        \left(\displaystyle{\d\sigma(\rho)\over \d\Omega}
                -\displaystyle{\d\sigma(\tilde \rho)\over \d\Omega}\right)
\right]^2
\over
\displaystyle\int\!\d\Omega\ \displaystyle{\d\sigma(\tilde \rho)\over \d\Omega} }
~.\label{e4}
\eea
The probability that a measurement of $\rho$ 
deviates from $\tilde \rho$
is quantified by $\chi_1^2$:
the computed interval of $\rho$
for which $\chi_1^2$ is less than a certain number
(say 2.71)
will contain a measured value of $\rho$ 
with the corresponding confidence level
(here 90\%).
The size of this interval 
is the precision with which the parameter can be determined
by measuring the total \xs.

The extent of this error band around $\tilde \rho$
depends of course on the value of $\tilde \rho$.
If experimental data is available,
$\tilde \rho$ is taken to be the best fit of $\rho$ to this data.
In the absence of actual data\footnote{
This is the situation we consider from now on.
},
though,
the value of $\tilde \rho$ is the result of an educated guess
or a theoretical bias,
typically, the \sm\ expectation.

Up to now
only a very small portion of the available information
has been used.
Indeed,
it might well be that 
two very different values of $\rho$
yield the same number of events.
Still,
these events might have significantly different topologies.
Upon integrating over the whole phase space in Eq.~(\ref{e4}),
these differences are completely washed out.
Striking examples of this phenomenon
have been discussed in Refs~\cite{df}.

Clearly,
it would be advantageous to include at least some of the information 
contained in the event shape.
This is usually done by considering asymmetries
or by dividing the phase space into a certain number $N$ of intervals
of one or several kinematical variables 
$\Delta\Omega_i$ ($i=1\dots N$).
The previous least squares estimator can then be applied separately
to each bin in these kinematical variables:
\beq
\chi^2_N = \sum_{i=1}^N \left( {n_i(\rho)-n_i(\tilde \rho) 
\over \Delta n_i(\tilde \rho)} \right)^2
~,\label{e5}
\eeq
where the index $i$ denotes a particular phase space bin
and $N$ is the total number of bins.
This is a standard procedure which can substantially improve
the resolving power of an experiment
\cite{adf}.
Indeed,
because of the triangle inequality
$\chi^2_N$ can only grow with the number of bins $N$
and one always has 
$\chi^2_N \ge \chi^2_1$.

Of course,
strictly speaking 
the quantitative probabilistic interpretation of this analysis is only valid 
as long as the number of bins is not excessive
and each bin contains a certain minimum number of events,
typically five.
Indeed,
a $\chi^2$ distribution is defined to be 
the weighted sum of the squares of independent gaussian distributions.
However,
if too many too small bins are used,
this definition is not obeyed
for two reasons:
\begin{description}
\item[A]The binning of the final state phase space
        takes place with a certain instrumental error,
        which introduces some amount of bin-to-bin correlation.
        The numbers of events in different bins
        are thus not completely independent.
\item[B]The number of events in each bin is in reality
        distributed according to a Poisson distribution,
        which assumes only asymptotically a gaussian shape.
\end{description}
Obviously,
if the number of bins is taken to be so large 
that the calculated number of events in some bins is less than one,
the whole procedure stops making sense.

Notwithstanding this limitation,
let us increase
(at least on paper)
the number of bins to infinity!
In this limit
the number of events per bin
\beq
n_i = {\cal L} \int_{\Delta\Omega_i}\!\d\Omega\ {\d\sigma \over \d\Omega}
\approx {\cal L} \left.{\d\sigma \over \d\Omega}\right|_{\Omega_i} \Delta\Omega_i
\label{e6}
\eeq
is infinitesimally small
and $\chi^2_N$ (\ref{e5}) becomes
\beq
\chi^2_\infty =
{\cal L}\
\int\!\d\Omega\
{\left(
        \displaystyle{\d\sigma(\rho)\over \d\Omega}
        -\displaystyle{\d\sigma(\tilde \rho)\over \d\Omega}
\right)^2
\over
        \displaystyle{\d\sigma(\tilde \rho)\over \d\Omega} 
}
~.\label{e7}
\eeq
Comparing this with $\chi^2_1$ (\ref{e4}),
we see that in essence 
the square of an integral 
became the integral of a square.
Clearly
\beq
\chi^2_\infty \ge \chi^2_N \ge \chi^2_1
~,\label{e8}
\eeq
so $\chi^2_\infty$ is the most sensitive estimator of $\rho$.

Because in some sense we assumed an infinite data sample
when taking the limit (\ref{e6}),
this is the asymptotic resolution
which could also be obtained by the \ml\ method.
Indeed,
defining the probability density
\beq
p = {1\over\sigma} {\d\sigma\over\d\Omega}
~,\label{e81}
\eeq
when $\rho$ is in the neighbourhood of $\tilde \rho$,
$\chi^2_\infty$ (\ref{e7}) can be rewritten 
in the linear regime\footnote{
        {\em i.e.} either if the dependence of $p(\rho)$ 
        on the parameter $\rho$ is linear
        or if the considered values of $\rho$ are close enough to $\tilde \rho$
        to warrant sufficient linearity}
as
\bea
\chi^2_\infty 
&=&
n \left(\rho-\tilde \rho\right)^2 \int\!\d\Omega\ {1\over p}
\left.\left({\partial p \over \partial \rho}\right)^2 \right|_{\tilde \rho}
\nonumber\\
&=&
\left(\rho-\tilde \rho\right)^2 \left<\left.-{\partial^2\ln L 
\over \partial \rho^2}\right|_{\tilde \rho}\right>
~,\label{e82}
\eea
which is nothing but the \ml\ estimator \cite{pdg},
where 
\beq
L = \prod_{i=1}^n\ p(\Omega_i)
\label{e83}
\eeq
is the \ml\ function.

To see that this is indeed the Cram\'er-Rao minimum variance bound,
we set $\chi^2_\infty=1$ in Eq.~(\ref{e82}).
Discretizing again into phase space bins,
we obtain for the dispersion of $\rho$ around $\tilde \rho$
\beq
D(\rho)^{-1}
=
\left.{1\over\left(\rho-\tilde \rho\right)^2}\right|_{\chi^2_\infty=1}
=
\sum_i {1\over n_i} \left({\partial n_i \over \partial \rho}\right)^2
~.\label{e91}
\eeq
By definition,
$n_i$ is the average number of events in bin $i$.
The observed number of events $N_i$ in this bin 
is distributed according to Poisson statistics,
{\em i.e.},
\beq
p_i = {e^{-n_i} n_i^{N_i} \over N_i!}
\label{e92}
\eeq
is the probability to find $N_i$ events in bin $i$.
Assuming there are no bin-to-bin correlations,
we have
\bea
<N_i> \quad&=&\quad n_i
\label{e93}
\\
<(N_i-n_i)(N_j-n_j)> \quad&=&\quad \delta_{ij} n_i
\label{e94}
\eea
and we can rewrite
\bea
D(\rho)^{-1}
&=&
\sum_{i,j} \left< \left({N_i\over n_i}-1\right) \left({N_j\over n_j}-1\right) \right>
{\partial n_i\over\partial \rho} {\partial n_j\over\partial \rho} 
\nonumber\\
&=&
\left<\left( \sum_i \left({N_i\over n_i}-1\right) {\partial n_i\over\partial \rho} \right)^2\right>
~.\label{e95}
\eea
This is nothing 
but the Cram\'er-Rao minimum variance bound\footnote{
  I am indebted to Sergey Alekhin
  for pointing out this derivation to me.}
\beq
D(\rho)^{-1}
=
\left<\left( \sum_i {\partial \ln p_i\over\partial \rho} \right)^2\right>
~.\label{e96}
\eeq

To derive this result,
we only assumed the absence of bin-to-bin correlations 
in Eq.~(\ref{e94}).
No assumption concerning the population of the bins is necessary.
Although we used the linear approximation
in Eq.~(\ref{e91}),
Eq.~(\ref{e7}) remains valid
even when the parameter dependence is far from linear,
which is often the case
when the luminosity $\cal L$ is small.
In contrast, 
the relations (\ref{e82}) assume a linear parameter dependence
because they are derived from the \ml\ covariance matrix.

In the presence of real data
the \ml\ function (\ref{e83}) can easily be evaluated
with all experimental resolutions and efficiencies folded in \cite{tim}.
The linear approximation is then not any longer necessary
since the confidence intervals can be estimated 
without having recourse to the covariance matrix.
In contrast,
the $\chi^2_\infty$ estimator can of course not be applied experimentally,
since it assumes
({\bf A}) the absence of systematical errors and
({\bf B}) sufficient statistics to fill infinitesimal bins.
These limitations, however,
only emphasize the fact that $\chi^2_\infty$
yields the theoretical Cram\'er-Rao limit
of what can be measured by the reaction.
In other words,
any data analysis of a particular reaction, 
however clever, 
cannot yield a more precise determination of a given parameter
than the asymptotic accuracy yielded by the $\chi^2_\infty$ estimator.

If the systematic errors
can be neglected with respect to the statistical error,
the Cram\'er-Rao bound predicted by the $\chi^2_\infty$ estimator (\ref{e7})
can be experimentally reached 
with a \ml\ analysis.
However,
if the systematic errors are large,
the question arizes,
how close can one come in practice 
to the theoretical precision 
given by the $\chi^2_\infty$ estimator?
There is no general answer to this question
and a separate analysis has to be performed for each case.
This issue is addressed in the next Section
at hand of a simple example.

\section{Electric Dipole Moment of the Electron in M\o ller Scattering}

To illustrate how the $\chi^2_\infty$ estimator works in practice,
let us analyze a particularly simple example.
If the electron is a composite particle,
its non-elementary nature might reveal itself 
at energies far below it's binding energy
by an electric dipole moment $d$.
This dipole plays now the role of the parameter $\rho$.
The electron-photon coupling is then described by the effective lagrangian
\beq
{\cal L} = -ie\bar\psi\gamma^\mu\psi A_\mu
           -i{d\over2}\bar\psi\sigma^{\mu\nu}\gamma_5\psi F_{\mu\nu}
~,\label{e101}
\eeq
where $e$ and $d$ are the \EM\ charge and electric dipole moment 
of the electron,
$F_{\mu\nu}=\partial_\mu A_\nu-\partial_\nu A_\mu$
is the strength of the \EM\ field $A_\mu$
and 
$\sigma^{\mu\nu}=\left(\gamma^\mu\gamma^\nu-\gamma^\nu\gamma^\mu\right)$.
The first term in the lagrangian (\ref{e101})
represents the {\em standard} point-like electron-photon coupling,
whereas the second term arizes from new interactions.

The static limit for such an electric dipole of the electron
is very tightly constrained by low energy experiments \cite{berk}.
However,
such a dipole term might well assume large values
for high momentum transfers \cite{leo},
if it behaves as a function of the photon virtuality $Q^2$ as
\beq
d \sim {Q\over\Lambda^2}
~,\label{e100}
\eeq
where $\Lambda$ is the scale of new physics.

To probe this electric dipole moment of the electron,
let us consider a polarized M\o ller scattering experiment.
It has the virtue of being particularly simple
and to allow the description of some important features 
of the $\chi^2_\infty$ estimator
with handy analytic formulae.
The $e^-e^-$ reaction takes place at lowest order in perturbation 
via the $t$- and $u$-channel exchanges 
of a photon or a neutral vector boson $Z^0$.
In the absence of transverse polarization
the final state phase space is one-dimensional.
Neglecting the mass of the electron and terms of ${\cal O}(d^4)$,
the differential \xs\ for {\em left-polarized} electron beams becomes
\beq
{\d\sigma \over \d\cos\theta}
=
{e^4 \over \pi s}
{1 \over \sin^4\theta}
\left(
        1 + {d^2s \over 2e^2} \sin^2\theta \cos^2\theta
\right)
~,\label{e102}
\eeq
where $\theta$ is the polar angle of the emerging electrons
and $\sqrt{s}$ is the \cm\ energy.
To derive Eq.~(\ref{e102})
we have ignored the $Z^0$ exchange.
This approximation doesn't introduce any qualitative change,
but has the virtue of keeping the analytic expressions simple.
In our numerical calculations
the $Z^0$ is of course taken into account.

Such a M\o ller scattering experiment
will be possible at one of the \lc s of the next generation
(CLIC, JLC, NLC, TESLA,\dots).
To be specific,
we concentrate here on the canonical design 
with a \cm\ energy $\sqrt{s}=500$ GeV
and an integrated luminosity ${\cal L}=10$ fb$^{-1}$.
In practice,
also,
the scattered electrons 
can only be observed at a certain angle away from the beampipe.
We therefore impose the angular cut
\beq
\cos\theta < 1-\epsilon
~.\label{e103}
\eeq

\begin{figure}
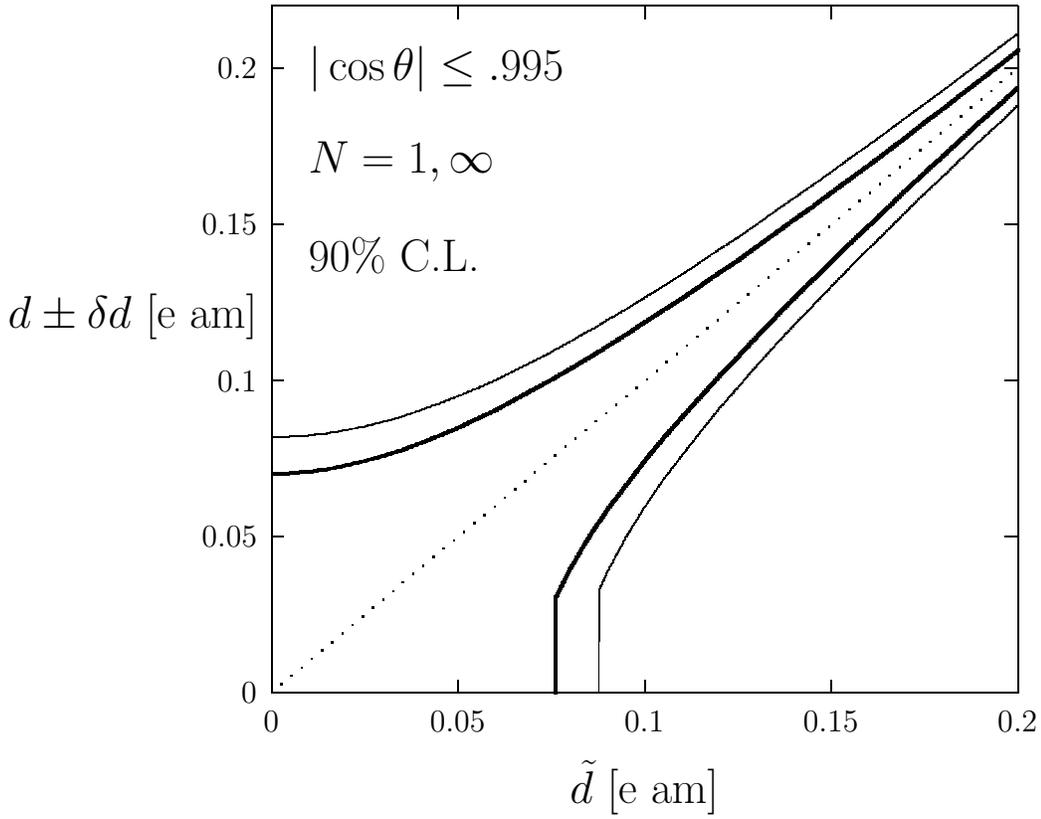

\input dev.tex
\caption{
Dependence of the resolving power
on the actual value of the electric dipole moment of the electron
$\tilde d$.
The resolution with one bin
(total \xs)
and an infinite number of bins
(the Cram\'er-Rao limit)
are given by the thinner and thicker curves respectively.
}
\label{dev}
\end{figure}

Of course,
the resolving power of this reaction
depends on the true value $\tilde d$ of the parameter.
In Fig.~\ref{dev} 
the 90\%\ confidence level
error band around $\tilde d$ 
(derived from $\chi^2_1,\chi^2_\infty=2.71$) 
is plotted as a function of $\tilde d$.
Since only $|d|^2$ can be observed in this experiment,
the plot extends in the same way in the three other quadrants.
For (not too) large values of $\tilde d$
the resolution scales like
$$
{1 \over \tilde d} \sqrt{{\chi^2 \over {\cal L}}}~.
$$
Indeed,
the expression for $\chi^2_1$ 
approaches in the limit of a vanishing cut $\epsilon$
\beq
\chi^2_1 \simeq {{\cal L}s\over\pi} \left(d^2-\tilde d^2\right)^2 2\epsilon
~.\label{e104}
\eeq
The reason why $\chi_1^2$ has no sensitivity 
when the whole kinematical range is inspected
($\epsilon\to0$),
can be traced back to the fact that the dipole moment
induces no singularity along the beampipe,
in contrast to the point-like coupling.
If small angle electrons are also considered,
the \sm\ background keeps increasing
whereas the dipole signal does not improve.
The collinear divergence of the \sm\ \xs\ 
is eventually regulated by the mass of the electron.
Strictly speaking, thus,
$\chi^2_1$ in (\ref{e104}) converges to a {\em very} small 
but finite value.
For our purposes, though, 
this effect is of no importance.

\begin{figure}
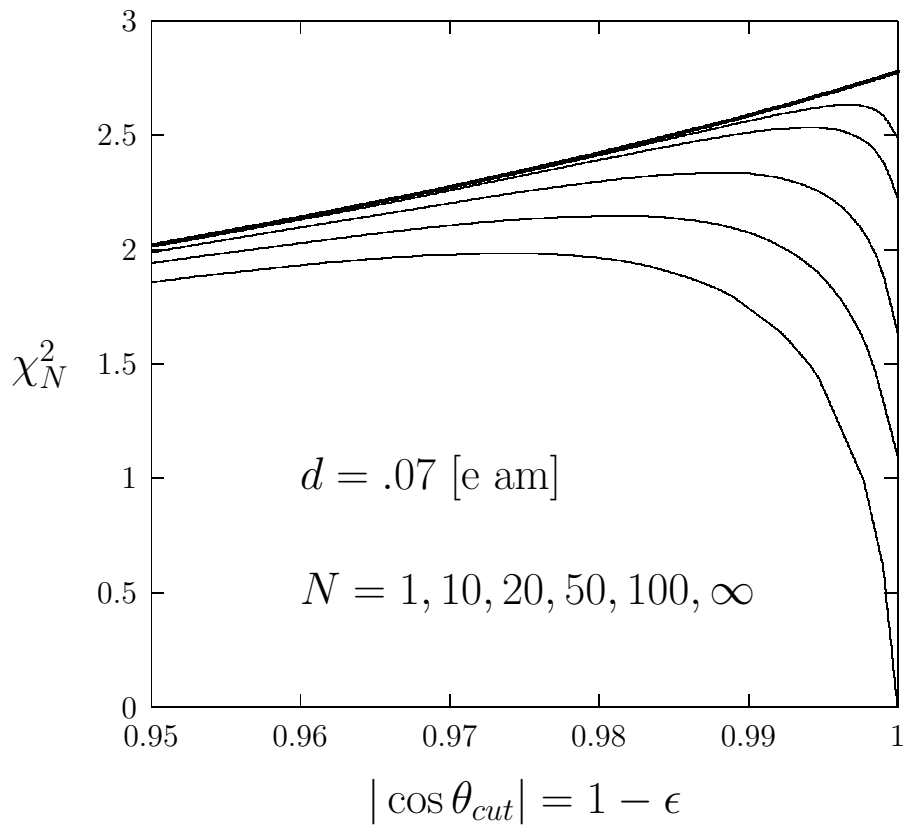

\input cut.tex
\caption{
Dependence of $\chi^2_N$ 
on the angular cut Eq.~(\protect\ref{e103}).}
\label{cut}
\end{figure}

The angular cut (\ref{e103}) could be optimized 
({\em cf.} Fig.~\ref{cut})
to maximize $\chi^2_1$
\cite{sasha}.
However, 
a partition of the angular range into a reasonable number $N$ of bins
automatically takes care of this task.
For the asymptotic limit
we find the approximate result
\beq
\chi^2_\infty \simeq {{\cal L}s\over\pi} \left(d^2-\tilde d^2\right)^2 
{(1-\epsilon)^5\over10}
~.\label{e105}
\eeq
This is the theoretical limit
which can only be approached from below 
by any experimental setup.

\begin{figure}
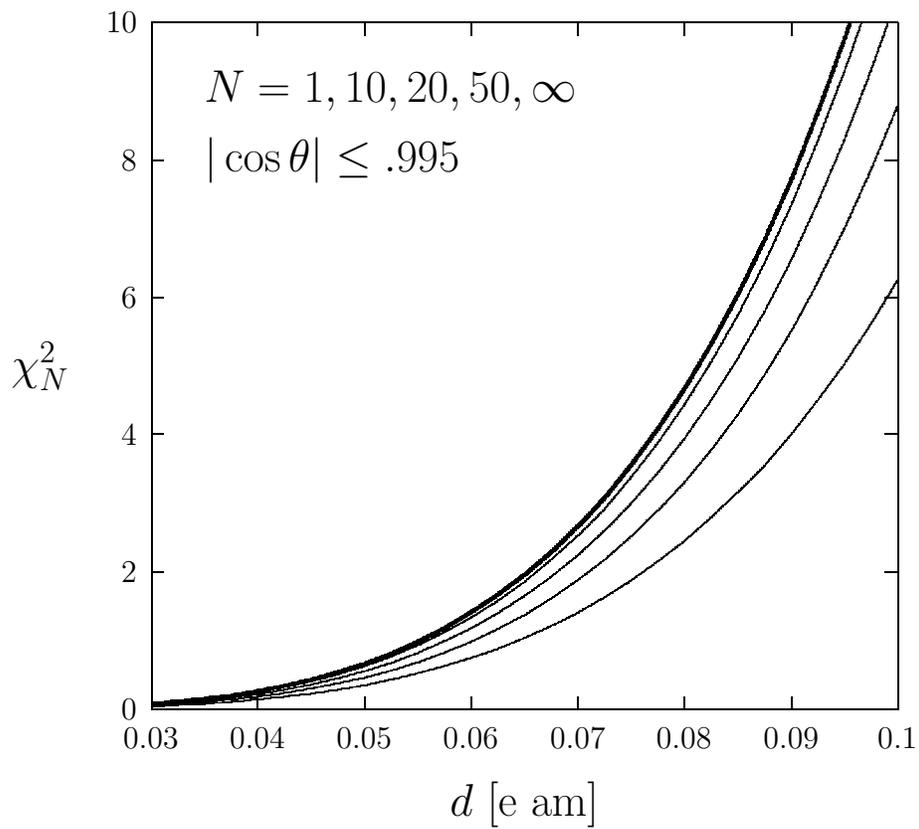

\input dip.tex
\caption{
Dependence of $\chi^2_N$ 
on the electric dipole moment of the electron $d$.}
\label{dip}
\end{figure}

To study the improvement of $\chi^2_N$
with increasing number of bins,
let us assume the validity of the \sm,
{\em i.e.},
$\tilde d=0$.
This way we test the limit of observability 
of the electron's electric dipole moment.
The deviations from $d=0$
which can be observed with a certain level of confidence
(say again 90\%)
are the values of $d$ 
which yield a $\chi^2$ in excess of a given number
(here again 2.71).
In Figs~\ref{dip} and \ref{cut}
the $d^4$ dependence
and the angular cut $\epsilon$ behaviours of $\chi^2_1$ and $\chi^2_\infty$
can be observed to agree with Eqs~(\ref{e104}) and (\ref{e105}).

\begin{figure}
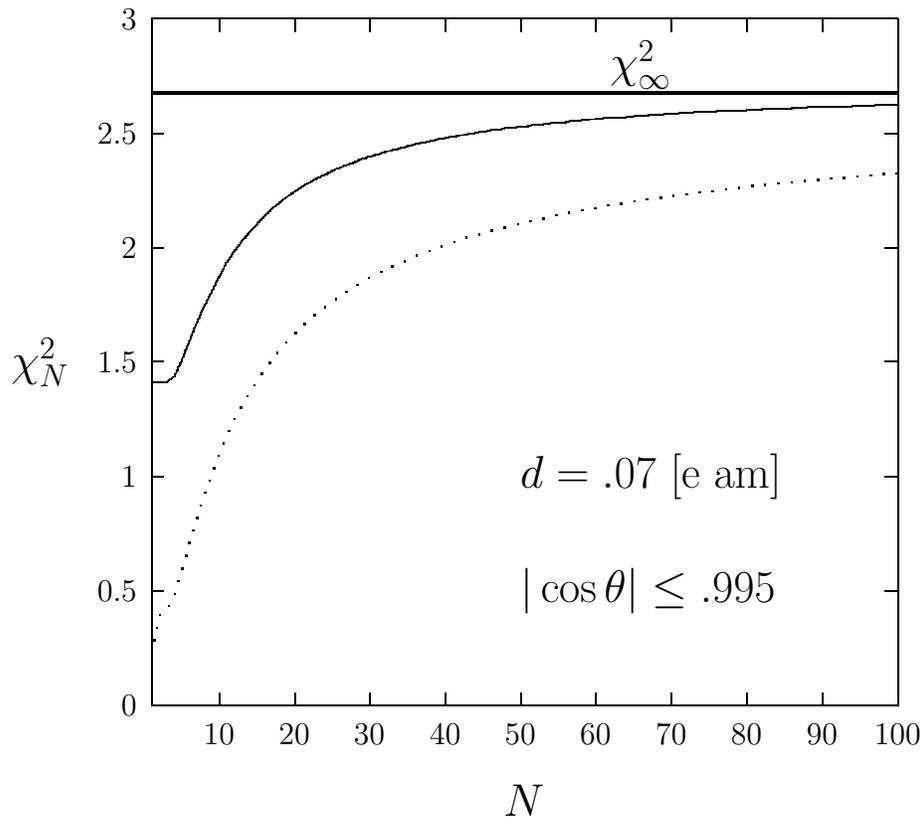

\input bin.tex
\caption{
Dependence of $\chi^2_N$ 
on the number of bins $N$.
The effect of systematic errors is shown by the dotted curve.}
\label{bin}
\end{figure}

It also appears from Fig.~\ref{bin},
where $\chi^2_N$ is plotted as a function of the number of bins $N$,
that with only 30 bins
one comes within 90\%\ of the asymptotic resolution.
Because the event rates of this reaction are so large,
however,
the error is in this case dominated by systematics.
Assuming for this very clean experiment a .1\%\ systematic error,
the expected results
are displayed by the dotted curve in Fig.~\ref{bin}.

\section{Conclusions}

We have presented a simple $\chi^2_\infty$ estimator
to evaluate the potential of a reaction
for studying parameters.
The estimator reveals the highest accuracy
this reaction could provide
under ideal conditions,
for determining the numerical values of these parameters:
the Cram\'er-Rao bound.

This estimator does not make any claim about the precision 
to be obtained under normal running conditions,
except that it can never be better.
In practice,
however,
this limit can be closely approached 
by a \ml\ data analysis,
if the systematic errors are not too large.

Since the $\chi^2_\infty$ estimator provides a bound 
on what precision can be achieved 
by a particular reaction
in the best of all cases,
it is a safe measure
to compute this number
before embarking on a more time consuming detailed analysis.
It can then be decided whether or not this reaction 
has at all a chance to compete in precision with others.

\begin{ack}
It is a pleasure to thank Sergey Alekhin, Tim Barklow and 
Geert Jan van Oldenborgh
for their enlightening comments and suggestions,
as well as Wolfgang Ochs, Stefan Pokorski and Roberto Pittau
for their critical comments on the readability of the manuscript.
\end{ack}

\clearpage

\clearpage

\clearpage

\clearpage

\begin{comment}{
\clearpage
\begin{figure}
\input rat.tex
\caption{Compton scattering:
dependence of the efficiency $\chi^2_N/\chi^2_\infty$ 
on the number of bins $N$.}
\label{rat}
\end{figure}
}\end{comment}

\end{document}